\documentclass[10pt,twocolumn,letterpaper]{article}

\usepackage[letterpaper,left=0.8125in,right=0.8125in,top=1in,bottom=1in,columnsep=0.3125in]{geometry}
\usepackage{times}
\usepackage{xspace}
\usepackage[dvipsnames]{xcolor}
\usepackage{graphicx}
\usepackage{amsmath}
\usepackage{amssymb}
\usepackage{booktabs}
\usepackage[numbers,sort&compress]{natbib}
\usepackage{etoolbox}
\usepackage[format=plain,labelformat=simple,labelsep=period,font=small,compatibility=false]{caption}
\usepackage[font=footnotesize,skip=3pt,subrefformat=parens]{subcaption}
\usepackage[hyphens]{url}
\usepackage[shortlabels,inline]{enumitem}
\setlist[itemize]{noitemsep,leftmargin=*,topsep=0em}
\setlist[enumerate]{noitemsep,leftmargin=*,topsep=0em}
\usepackage{authblk}

\usepackage{microtype}
\usepackage{mathtools}

\graphicspath{{figures/}}

\newcommand{\method}{H-2DGS\xspace}
\newcommand{\stageone}{Stage~I\xspace}
\newcommand{\stagetwo}{Stage~II\xspace}
\newcommand{\stagethree}{Stage~III\xspace}

\DeclareMathOperator{\clip}{clip}
\DeclareMathOperator{\SSIM}{SSIM}

\definecolor{linkblue}{rgb}{0.21,0.49,0.74}
\usepackage[breaklinks,colorlinks,allcolors=linkblue]{hyperref}
\usepackage[capitalize]{cleveref}
\crefname{section}{Sec.}{Secs.}
\Crefname{section}{Section}{Sections}
\Crefname{table}{Table}{Tables}
\crefname{table}{Tab.}{Tabs.}

\title{\method: Height-Field Gaussian Splatting for Sharp and Deformable Neurosurgical Surface Reconstruction}

\author[1]{\normalsize Zhang Zichuan}
\author[2]{\normalsize Bart Brouwers}
\author[1,2]{\normalsize Rembrandt Bakker}
\author[2]{\normalsize Maud Landers}
\author[1,2]{\normalsize Geert-Jan Rutten}
\author[1]{\normalsize Maxime Chamberland}
\affil[1]{\normalsize\textit{Department of Computer Science \& Mathematics, Eindhoven University of Technology, Eindhoven, The Netherlands}}
\affil[2]{\normalsize\textit{Department of Neurosurgery, Elisabeth-Tweesteden Hospital Tilburg, The Netherlands}}
\date{}

\begin{document}
\maketitle

\begin{abstract}
Neurosurgical navigation commonly presents magnetic resonance imaging (MRI) as 2D slices, making it difficult to relate the anatomy in those slices to the cortical surface observed through the surgical microscope.
We seek a common 3D visualization that combines a photorealistic reconstruction of the exposed cortex during tumor surgery with whole-brain MRI context and diffusion tractography pathways.
This requires the reconstructed cortex to form a clear boundary around the brain anatomy rather than a diffuse volume that obscures internal structures.
Conventional 3D Gaussian Splatting (3DGS) can reproduce observed views but, under limited surgical view coverage, may distribute opacity through a thick and geometrically ambiguous band.
We introduce height-field 2D Gaussian Splatting (\method), a three-stage pipeline that concentrates a robust 3DGS reconstruction, fuses its depth into a shared truncated signed distance field (TSDF) surface, and represents that surface with curved, surface-oriented Gaussian primitives.
Separate height, color, and opacity bases provide local geometry and appearance, while a neighbor-transition objective coordinates overlapping patches.
The explicit surface further supports postoperative alignment: MRI-derived depth deforms the reconstruction, and tracked-color supervision with alternating optimization preserves its appearance.
Across five real neurosurgical cases, \method gives low multi-view depth inconsistency, a compact 7.97\,MB representation, and interactive rendering at 29--42\,ms per frame.
The resulting sharp, deformable surface integrates clearly with tractography and MRI anatomy, including a postoperative resection cavity. 
\end{abstract}

\begin{figure*}[t]
    \centering
    \begin{subfigure}[t]{0.48\textwidth}
        \centering
        \includegraphics[width=\linewidth]{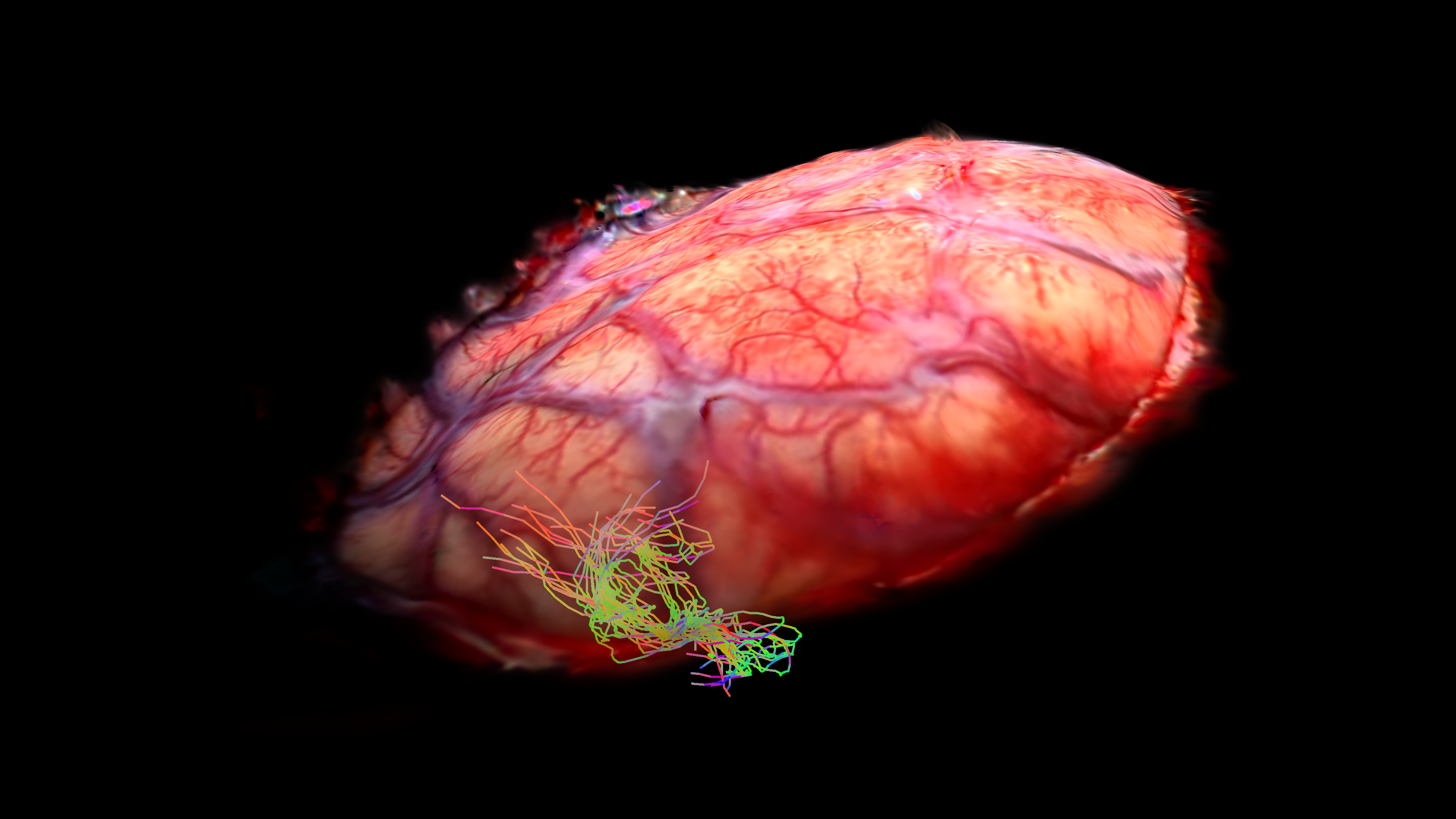}
        \caption{Fusion of reconstructed cortex and tractography (streamlines) derived from diffusion MRI data.}
        \label{fig:teaser-tracts}
    \end{subfigure}
    \hfill
    \begin{subfigure}[t]{0.50\textwidth}
        \centering
        \includegraphics[width=\linewidth]{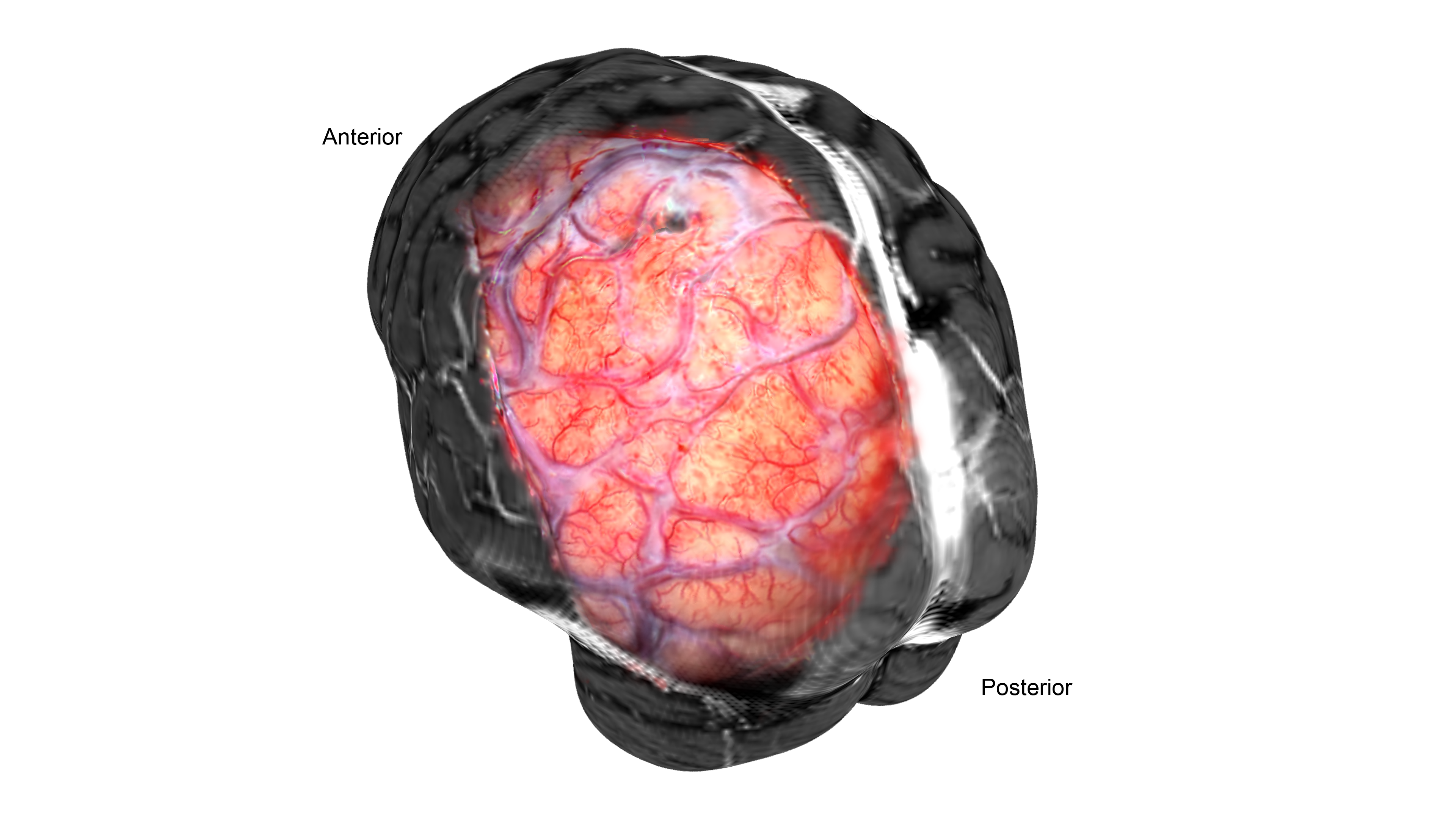}
        \caption{Pre-operative fusion with an MRI-derived surface (T1-weighted contrast-enhanced).}
        \label{fig:teaser-postop}
    \end{subfigure}
    \caption{\textbf{A geometrically sharp, deformable surgical surface.}
    (a) \method reconstructs a localized photorealistic cortical layer that remains clear when combined with subcortical fiber tracts.
    (b) Preoperative alignment between the deformed surgical surface and MRI.}
    \label{fig:teaser}
\end{figure*}

\section{Introduction}
\label{sec:introduction}
Medical visualization plays a central role in the interpretation, communication and education of neuroanatomy and tractography.
Interactive visualization is important in, e.g., neurosurgical planning, as it enables users to explore anatomical structures and fiber pathways. Modern neurosurgery uses dedicated navigation systems for surgical planning, intraoperative localization, and postoperative assessment.
These systems commonly present registered magnetic resonance imaging (MRI) as orthogonal 2D slices~\cite{example2dcombinedviews,BackgroundBrainShift}.
The slices contain rich tissue information, but they do not resemble the exposed cortical surface seen through the surgical microscope.
Consequently, understanding where a visible vessel, sulcus, or resection boundary lies within the MRI volume requires a mental correspondence between two very different views.
This separation increases cognitive load, and the missing 3D spatial context makes the relationship less intuitive~\cite{why3dgood}.

We instead aim to place the complementary information in one 3D scene.
Microscope video supplies the current color and fine surface landmarks of the exposed cortex.
MRI supplies the surrounding whole-brain tissue context, while diffusion MRI tractography exposes subcortical pathways that are not visible at the surface.
A photorealistic 3D reconstruction can connect what the surgeon sees to these internal structures under continuous viewpoint changes: the cortex provides the visible outer layer, the MRI-derived model locates that layer within the brain, and tractography shows the structures beneath it.
The fused views in \cref{fig:teaser} illustrate this intended relationship.

Creating this visualization is not only a registration problem; it first requires an appropriate representation of the visible layer.
3D Gaussian Splatting (3DGS)~\cite{Background3DGS} is attractive because it reconstructs realistic appearance, renders rapidly, and exposes primitives that can be manipulated.
It has consequently been adopted for static and deforming surgical scenes~\cite{endo-4dgx,endo-4dgs,free-dygs,free-surgs,endogaussian,deform3dgs}.
However, each conventional 3D Gaussian distributes opacity through a volume.
With the narrow camera trajectories typical of microscope video, Gaussians at several depths can explain the same training pixels.
The model may therefore look realistic from observed views while forming a thick, blurry, or fragmented density band elsewhere.
When tractography is rendered beneath this band, fibers can be obscured by or visually embedded in the reconstructed cortex, so the boundary between visible surface and subcortical anatomy becomes ambiguous.

We address this representation gap with \method, which progressively converts a robust volumetric reconstruction into a geometrically sharp yet photorealistic surface.
\stageone first uses conventional 3DGS to establish appearance without prematurely imposing a fragile surface.
\stagetwo then measures the spread of density along camera rays and contracts it toward the visible boundary.
Multi-view depth from this concentrated model is fused into a truncated signed distance field (TSDF), providing a shared surface rather than unrelated per-view estimates.
\stagethree discards the volumetric Gaussians and places curved height-field primitives on the TSDF zero level set.
The height field lets each patch follow local cortical curvature, separate color and opacity bases retain spatially varying texture, and neighbor-transition optimization coordinates overlapping patches.

The common 3D view must also remain meaningful when anatomy changes.
Brain motion and tissue removal alter the relationship between the reconstructed surface and MRI~\cite{BackgroundBrainShift}; after resection, the cavity can also invalidate vessel-based correspondence.
For the postoperative case, we therefore render depth from the MRI-derived mesh and deform the explicit height-field surface toward that geometry.
Depth determines the target shape, while tracked color supervision, overlap regularization, and alternating geometry--appearance updates prevent the alignment from erasing the photorealistic cortical texture.
Preoperative vessel-guided alignment is retained only as a downstream use of the reconstructed surface.

Our contributions are\footnote{Open source code will be provided via Github upon publication.}:
\begin{itemize}
    \item a three-stage conversion from volumetric 3DGS to a thin TSDF-initialized cortical surface, including a differentiable ray-wise thickness objective;
    \item a curved height-field Gaussian primitive with spatially varying geometry, color, and opacity, together with overlap constraints for locally continuous geometry;
    \item a depth-supervised deformation procedure that aligns the photorealistic surface to postoperative MRI geometry while preserving appearance, enabling unified 3D visualization with whole-brain anatomy and tractography.
\end{itemize}

\section{Related Work}
\label{sec:related}

\paragraph{Gaussian surface representations.}
Standard 3DGS represents a scene with anisotropic volumetric Gaussians optimized through differentiable alpha compositing~\cite{Background3DGS}.
Several alternatives strengthen its geometric prior.
2D Gaussian Splatting (2DGS) replaces ellipsoids by oriented disks and uses ray--splat intersection to obtain a more surface-like representation~\cite{2DGS}.
3D Half-Gaussian Splatting (3D-HGS) splits primitives with learned planes to sharpen discontinuities~\cite{3d-hgs}, while Quadratic Gaussian Splatting (QGS) models local geometry with second-order surfaces~\cite{QGS}.
These approaches initialize their designed primitives directly from an input point cloud.
Under limited view coverage, the source geometry can itself be weakly constrained.
\method instead first concentrates a robust volumetric reconstruction, fuses its multi-view depth into a shared TSDF, and only then initializes curved surface charts.
Per-primitive appearance bases are related to textured Gaussian representations~\cite{texgs1,texgs2,texgs3,texgs4}, but our height basis additionally makes each primitive a deformable local surface.

\paragraph{Surgical reconstruction and anatomical fusion.}
Neural and Gaussian representations have been applied to static and deforming surgical scenes~\cite{nerfsingleimage,endogaussian,deform3dgs}.
Most of these methods prioritize view synthesis or temporal tissue motion.
Cross-modal methods connect intraoperative imagery to preoperative anatomy through inverse rendering or coupled meshes~\cite{example3dcombinedviews2,bridgesplat}.
Our focus is complementary: we construct a sharp, compact cortical boundary from limited microscope views, use it as a substrate for tractography, and directly deform this renderable surface toward postoperative geometry.

\section{Sharp Height-Field Surface Reconstruction}
\label{sec:reconstruction}

\subsection{Problem formulation and design rationale}
\label{sec:pipeline}

The input is a set of calibrated microscope images and camera projections
\begin{equation}
    \mathcal{D}=\{(I_v,\Pi_v)\}_{v=1}^{V}.
    \label{eq:dataset}
\end{equation}
Our output is a renderable set of surface patches that must satisfy three requirements.
First, it should reproduce cortical color and vessel landmarks from the input views.
Second, opacity should be confined to a well-localized surface so that MRI anatomy and tractography remain visible on the correct side of the cortical boundary.
Third, the representation should expose depth and local geometric parameters for subsequent alignment.

Directly fitting surface-oriented splats from the sparse input point cloud is brittle in this setting: the same limited camera coverage that creates 3DGS thickness also leaves the initial geometry weakly constrained.
We therefore separate appearance acquisition from surface construction.
\stageone obtains a robust photometric reconstruction with flexible volumetric Gaussians.
\stagetwo makes this reconstruction a reliable geometric proxy by contracting its density along observed rays without changing the primitive type.
\stagethree fuses the resulting multi-view depths into one surface and replaces the volumetric model with curved height-field patches.
Thus each stage has a distinct purpose: bootstrap appearance, localize geometry, and encode an explicit surface.

\subsection{Stages I--II: from appearance to localized geometry}

\paragraph{Photometric bootstrap.}
\stageone trains conventional 3DGS with adaptive densification.
Starting from this flexible representation avoids forcing an inaccurate initial point cloud directly into a surface.
The model is optimized with the standard photometric objective
\begin{equation}
    \mathcal{L}_{\mathrm{photo}}
    =(1-\lambda)\|\hat I_v-I_v\|_1
    +\lambda\bigl(1-\SSIM(\hat I_v,I_v)\bigr).
    \label{eq:photo}
\end{equation}
to establish the visible tissue texture and overall scene support, where we set $\lambda$ = 0.2.
This objective determines how the training views should look, but it does not determine where along a camera ray the explaining opacity should lie.
The resulting model is therefore an appearance bootstrap rather than our final surface.

\paragraph{Ray-wise density concentration.}
The purpose of \stagetwo is to remove this depth ambiguity while retaining the successful image reconstruction.
A sharp visible boundary should generate a narrow density distribution along each foreground ray; a thick 3DGS band generates a broad or multimodal distribution.
We measure that spread with differentiable quantiles, which are less sensitive to isolated low-density tails than the extreme occupied samples.
For camera ray $\mathbf r(t)=\mathbf o+t\mathbf d$, substituting the ray into candidate 3D Gaussians gives one-dimensional contributions
\begin{equation}
    \sigma_i(t)=\beta_i\exp\!\left[-\tfrac12 a_i(t-m_i)^2\right],
    \qquad \sigma(t)=\sum_i\sigma_i(t).
    \label{eq:ray-density}
\end{equation}
At samples $\{t_m\}_{m=1}^{M}$, let
\begin{equation}
    p_m=\frac{\sigma(t_m)}{\sum_n\sigma(t_n)+\epsilon},
    \qquad F_m=\sum_{n=1}^{m}p_n .
    \label{eq:ray-cdf}
\end{equation}
We approximate quantile $q$ by
\begin{equation}
    w_m(q)=
    \frac{\exp(-\tau_q|F_m-q|)}
         {\sum_n\exp(-\tau_q|F_n-q|)},
    \qquad t_q=\sum_mw_m(q)t_m .
    \label{eq:soft-quantile}
\end{equation}
The effective thickness is $\Delta t=t_{q_{\mathrm{high}}}-t_{q_{\mathrm{low}}}$, producing
\begin{equation}
    \begin{aligned}
    \mathcal{L}_{\mathrm{thick}}(\mathbf r)
      &=\frac{\max(0,\Delta t-\Delta t^\star)}{E},\\
    \mathcal{L}_{\mathrm{stage2}}
      &=\mathcal{L}_{\mathrm{photo}}
        +\lambda_{\mathrm{thick}}\mathbb E_{\mathbf r}
          [\mathcal{L}_{\mathrm{thick}}].
    \end{aligned}
    \label{eq:thickness}
\end{equation}
where $E$ is the scene extent and $\lambda_{\mathrm{thick}}$ is set to 5.0.
We use $q_{\mathrm{low}}=0.01$, $q_{\mathrm{high}}=0.99$, and zero normalized target thickness.
\stagetwo runs for 1,000 iterations after a 5,000-iteration warm start.
The photometric term remains active, so contraction cannot simply collapse density at an arbitrary depth: it must preserve the observed cortical images while selecting a much thinner explanation.

\begin{figure}[!t]
    \centering
    \includegraphics[width=1.0\linewidth]{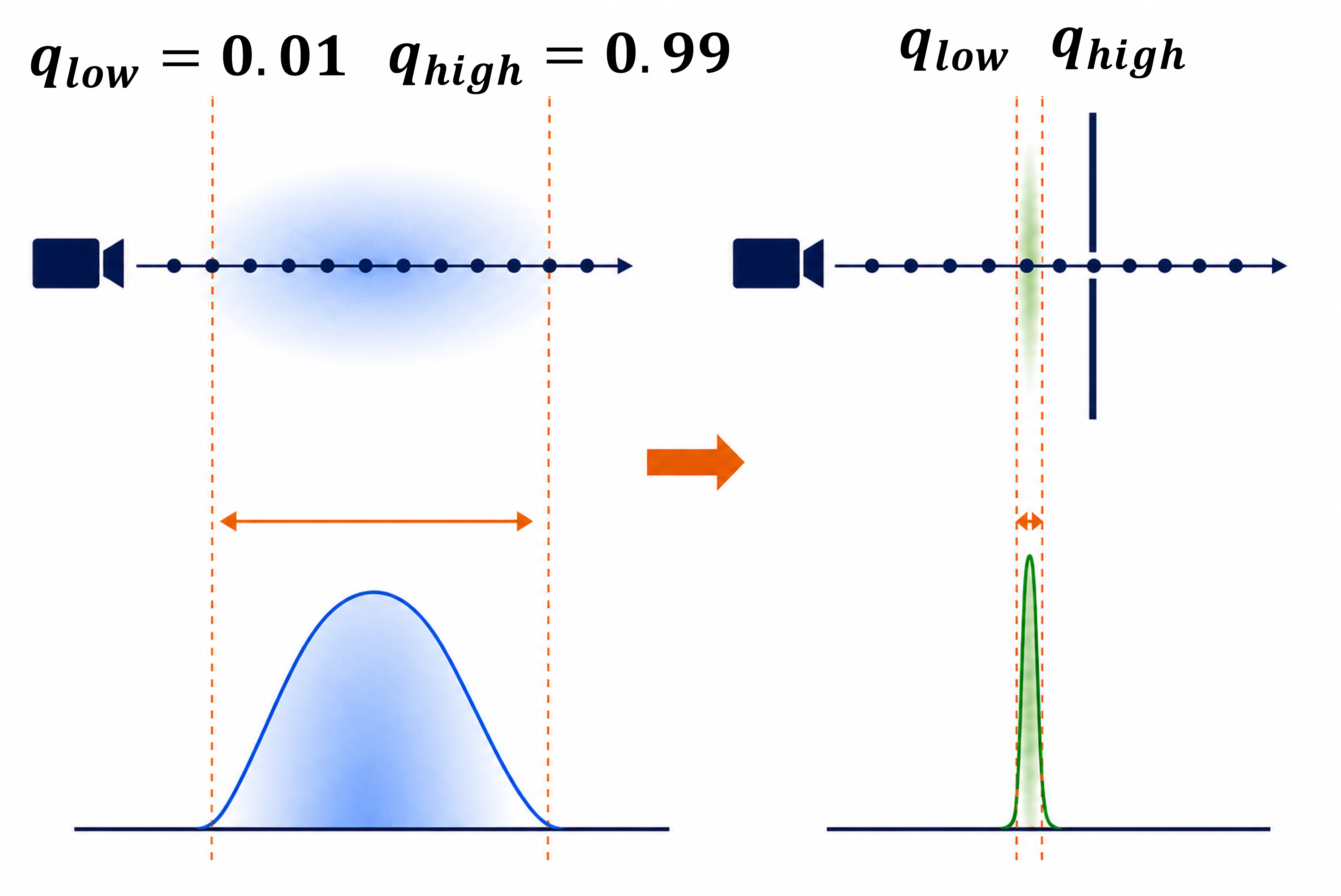}
    \caption{Ray-wise density concentration.
    \stagetwo contracts the density interval between soft 1\% and 99\% quantiles.}
    \label{fig:density}
\end{figure}

\subsection{Stage III: TSDF-guided surface conversion}

Although \stagetwo provides concentrated depth, it is still a collection of volumetric Gaussians and may contain residual floating density.
We therefore do not convert its individual ellipsoids directly into final primitives.
Instead, we render depth from all training cameras and fuse those observations into a truncated signed distance field (TSDF).
This intermediate field serves two purposes: it aggregates evidence across views into a single geometric reference, and its zero level set supplies both surface positions and normals for consistently orienting the new patches.
For voxel center $\mathbf x$, view $v$ compares its projected depth $d_v(\mathbf x)$ with the \stagetwo depth $\hat D_v(\pi_v(\mathbf x))$:
\begin{equation}
    \begin{aligned}
    \phi_v(\mathbf x)
      &=\clip\!\left(
        \frac{\hat D_v(\pi_v(\mathbf x))-d_v(\mathbf x)}{\mu},
        -1,1\right),\\
    \Phi(\mathbf x)
      &=\frac{\sum_vw_v(\mathbf x)\phi_v(\mathbf x)}
              {\sum_vw_v(\mathbf x)+\epsilon}.
    \end{aligned}
    \label{eq:tsdf}
\end{equation}
Only observations in the truncation band (we set $\mu$ to be three times of the voxel size) are fused.
Grid cells with a sign change provide zero-level candidates, which are projected toward $\Phi(\mathbf x)=0$.
The normalized gradient $\nabla\Phi/\|\nabla\Phi\|$ estimates surface normals.
After view-consistency filtering and subsampling, a local quadric fit supplies each patch center, tangent frame, curvature estimate, and support size.
The result is a sparse set of overlapping local charts that covers the shared TSDF surface rather than a noisy set of independently projected depth samples.

\begin{figure}[!t]
    \centering
    \includegraphics[width=1.0\linewidth]{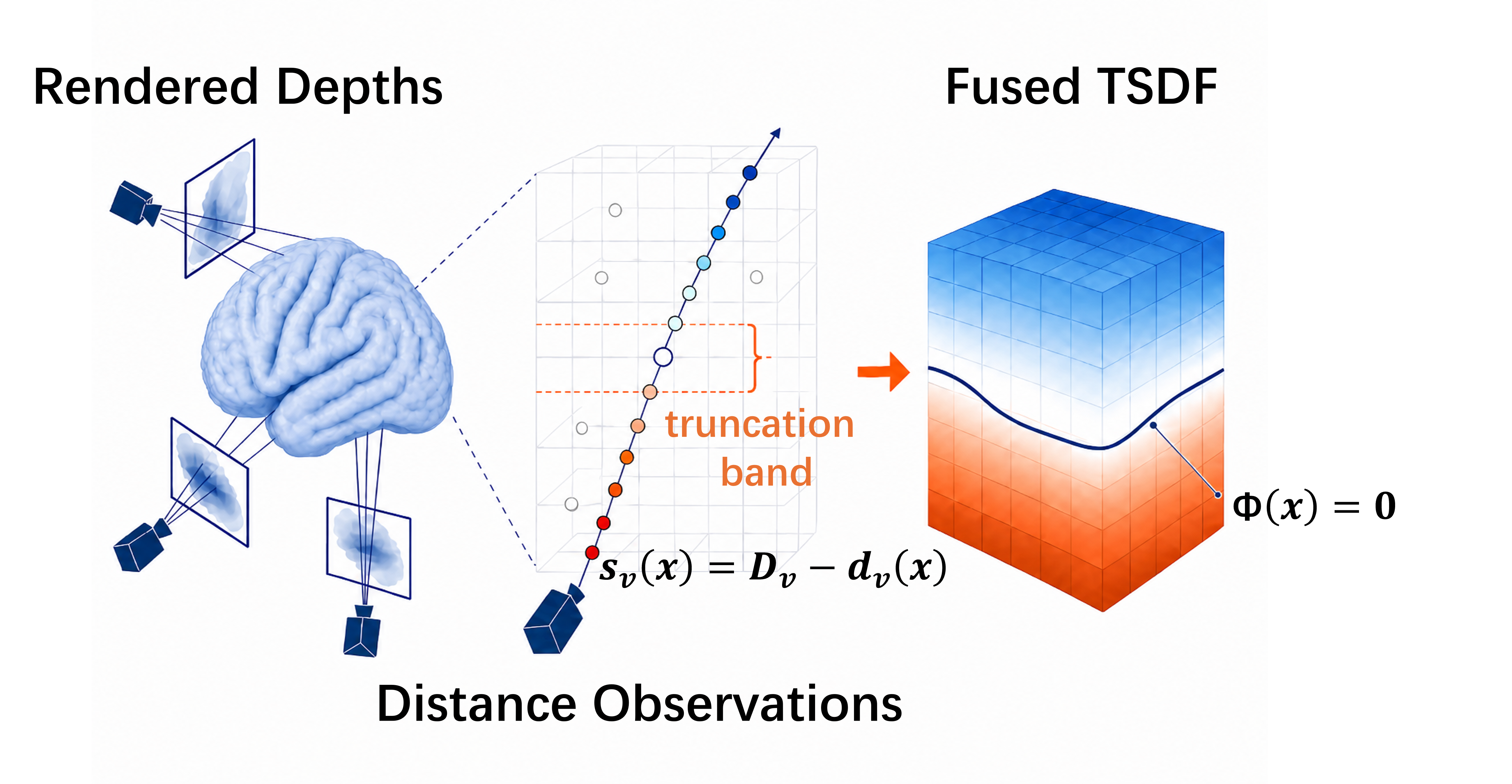}
    \caption{Multi-view TSDF fusion.
    Rendered depths define multi-view signed-distance observations whose fused zero level set initializes \stagethree.}
    \label{fig:tsdf}
\end{figure}

\subsection{Height-field Gaussian primitive}
\label{sec:primitive}

The TSDF could be represented with planar surfels, but a fixed plane can only approximate cortical curvature by using many small elements.
Our primitive keeps the thin support of a 2D surfel while allowing one patch to bend away from its tangent plane.
It also stores spatially varying appearance on that same local domain, so a relatively small number of patches can represent both smooth geometry and detailed cortical texture.
Primitive $i$ has center $\mathbf c_i$, tangent frame
$R_i=[\mathbf t_{u,i},\mathbf t_{v,i},\mathbf n_i]$, and tangent scales
$s_{x,i},s_{y,i}$.
Its surface is
\begin{equation}
    \mathbf S_i(u,v)=
    \mathbf c_i+
    R_i
    \begin{bmatrix}
        s_{x,i}u & s_{y,i}v & h_i(u,v)
    \end{bmatrix}^{\!\top},
    \label{eq:surface}
\end{equation}
with a Gaussian-basis height field
\begin{equation}
    \begin{aligned}
    h_i(u,v)
      &=\sum_{k=1}^{K_h}b^h_{i,k}\phi^h_{i,k}(u,v),\\
    \phi^h_{i,k}(u,v)
      &=\exp\!\left(
        -\frac{\|(u,v)-\boldsymbol\mu^h_{i,k}\|^2}
               {2(\sigma^h_{i,k})^2}\right).
    \end{aligned}
    \label{eq:height}
\end{equation}
Color $\mathbf C_i(u,v)$ and opacity $O_i(u,v)$ use analogous, independent bases plus constant terms. For each primitive, we use 5 height bases ($K_h$), 20 opacity bases and 50 color bases.
Geometry is intentionally lower-dimensional than appearance because surface shape varies more smoothly than cortical texture.
The color bases act as a local texture map, whereas the opacity bases control support and boundary strength inside the patch.
This separation lets appearance vary rapidly without introducing corresponding high-frequency oscillations in geometry.

\begin{figure}[!t]
    \centering
    \includegraphics[width=1.0\linewidth]{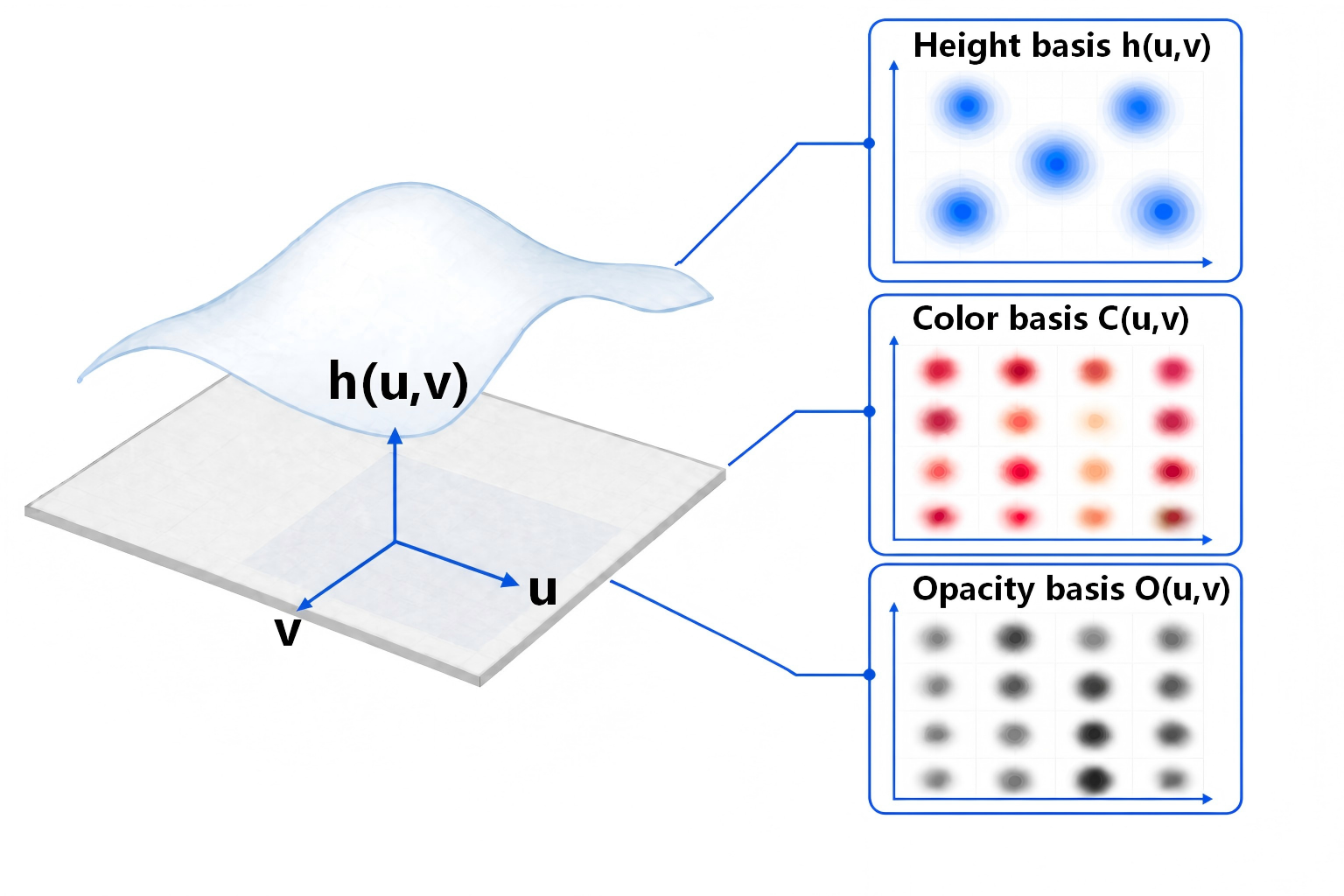}
    \caption{Concept of the proposed H-2DGS primitive.
    Each primitive is a curved local chart with independent basis fields over one tangent domain.}
    \label{fig:primitive}
\end{figure}

To render, a world-space ray is transformed as
$\mathbf y_i(t)=R_i^\top(\mathbf r(t)-\mathbf c_i)$.
Normalized coordinates are $u_i(t)=y_{x,i}(t)/s_{x,i}$ and
$v_i(t)=y_{y,i}(t)/s_{y,i}$.
The intersection is the root
\begin{equation}
    f_i(t)=y_{z,i}(t)-h_i(u_i(t),v_i(t))=0,
    \label{eq:intersection}
\end{equation}
found with a small fixed number of Newton iterations.
Color and opacity are evaluated at the accepted intersection and composited in depth order.
Unlike volumetric accumulation through an ellipsoid, this operation provides an explicit surface intersection and camera-linear depth, which are later reused by the postoperative alignment.

\subsection{Initialization, continuity, and refinement}

Surface positions alone are not enough to initialize the richer primitive.
For each patch, nearby TSDF zero-level samples are expressed in its tangent frame, and the height coefficients are fitted to their signed offsets from the tangent plane.
This converts the locally planar initialization into a curved chart before image refinement.
Nearby \stagetwo Gaussians transfer color and opacity to the new patch; the volumetric Gaussians are then discarded completely.

Because each chart is fitted locally, two individually accurate patches can still disagree where they overlap, producing cracks or abrupt normal changes.
We therefore sample corresponding locations in neighboring charts and coordinate them with
\begin{equation}
    \mathcal L_{\mathrm{trans}}=
    \lambda_{\mathrm{cont}}\mathcal L_{\mathrm{cont}}+
    \lambda_{\mathrm{pres}}\mathcal L_{\mathrm{pres}}+
    \lambda_{\mathrm{curv}}\mathcal L_{\mathrm{curv}}.
    \label{eq:transition}
\end{equation}
where $\lambda_{\mathrm{cont}}$ = 0.5, $\lambda_{\mathrm{pres}}$ = 2.0 and $\lambda_{\mathrm{curv}}$ = 0.02.
$\mathcal L_{\mathrm{cont}}$ penalizes world-space position and tangent mismatch at paired overlap samples;
$\mathcal L_{\mathrm{pres}}$ retains the TSDF-fitted heights and gradients; and
$\mathcal L_{\mathrm{curv}}$ suppresses oscillation.
The three terms have complementary roles: continuity closes patch interfaces, preservation prevents this pairwise adjustment from drifting away from the multi-view TSDF, and curvature regularization avoids fitting the correspondences with ripples.
After this geometric transition, final photometric refinement uses \cref{eq:photo} to recover cortical appearance on the already localized surface.
Shared basis layouts, precomputed inverse squared bandwidths, and CUDA-side indexing reduce rasterization overhead.

\begin{figure*}[t]
    \centering
    \includegraphics[width=1.0\textwidth]{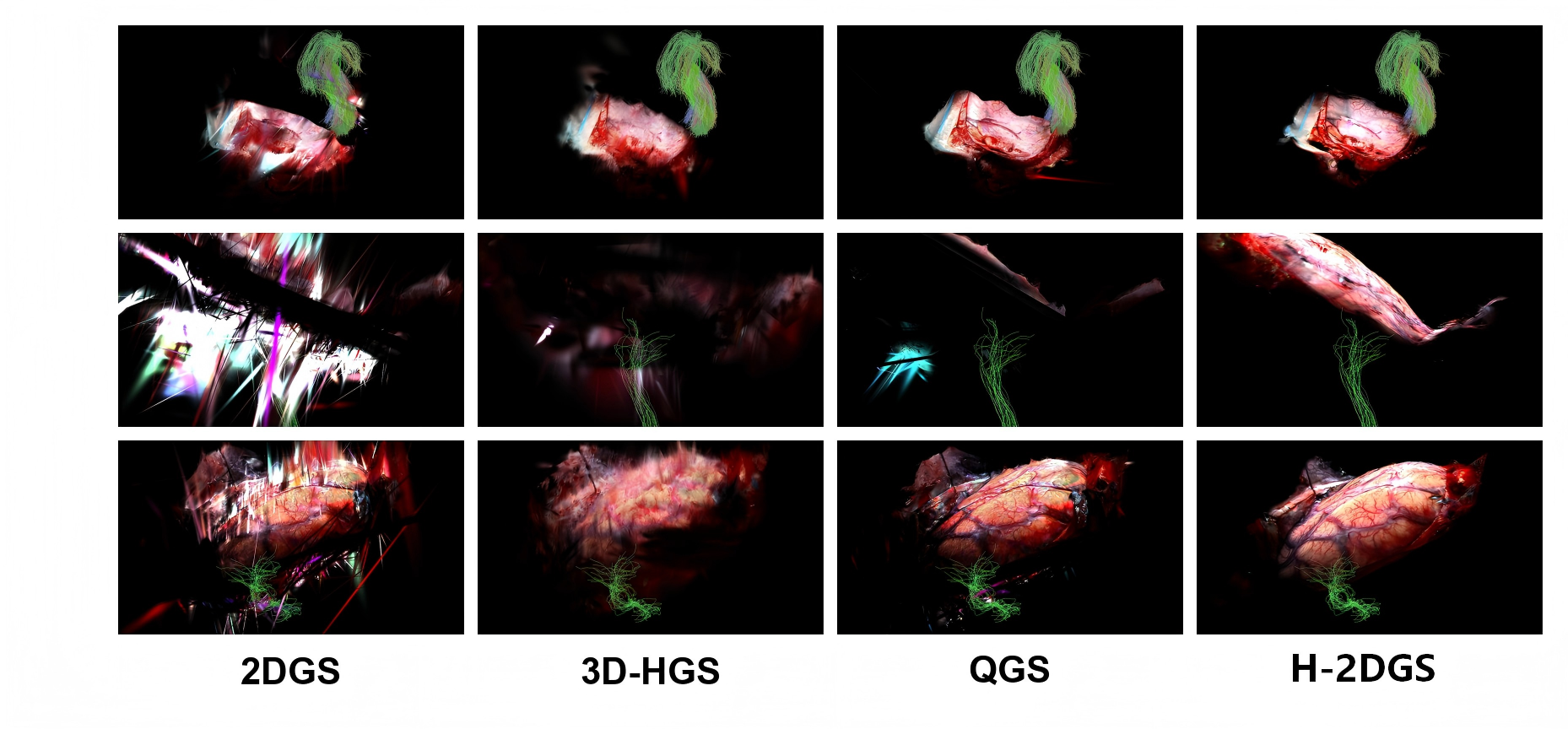}
    \caption{Fusion of reconstructed cortex and tractography derived from diffusion MRI. The top row shows the results of case preop 2, the second and the bottom rows show the results of case preop 3. columns are 2DGS, 3D-HGS, QGS, and \method.}
    \label{fig:fusion-comparison}
\end{figure*}

\begin{figure}[!t]
    \centering
    \includegraphics[width=1.0\linewidth]{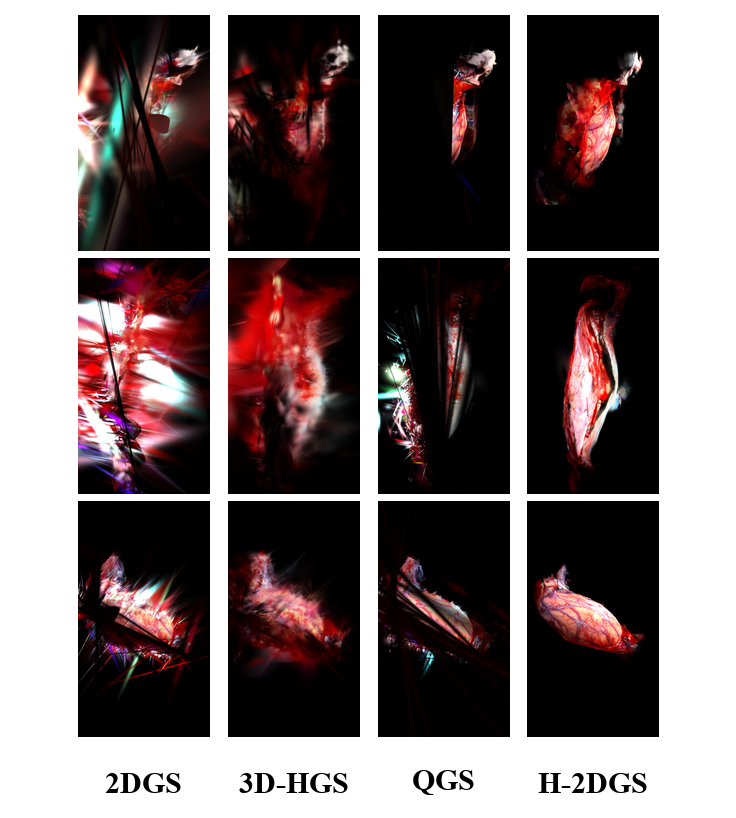}
    \caption{Uncovered-view comparison over three cases and four methods. Each row corresponds to the results of one of the four case. From top to bottom rows: results of case preop 1, preop 2 and preop 3 respectively.
    \method remains localized, whereas baseline geometry streaks, fragments, or blurs.}
    \label{fig:uncovered}
\end{figure}

\begin{figure}[!t]
    \centering
    \includegraphics[width=0.88\linewidth]{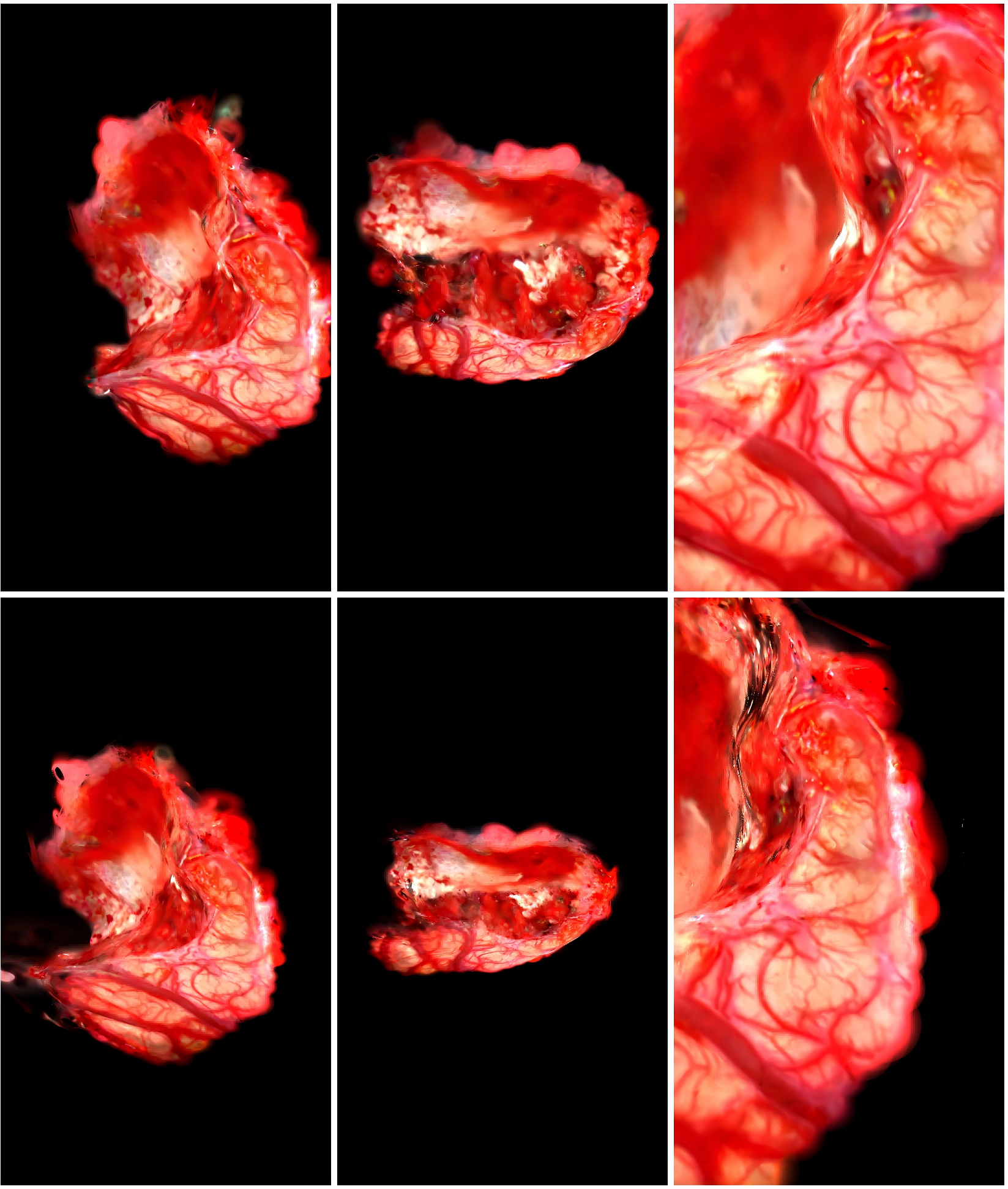}
    \caption{Result of deforming the post-resection reconstruction of the case postop 1. 
    Initial (top) and deformed (bottom) reconstruction fitted to postoperative MRI morphology.}
    \label{fig:postop-before-after}
\end{figure}

\begin{figure}[!t]
    \centering
    \includegraphics[width=1.0\linewidth]{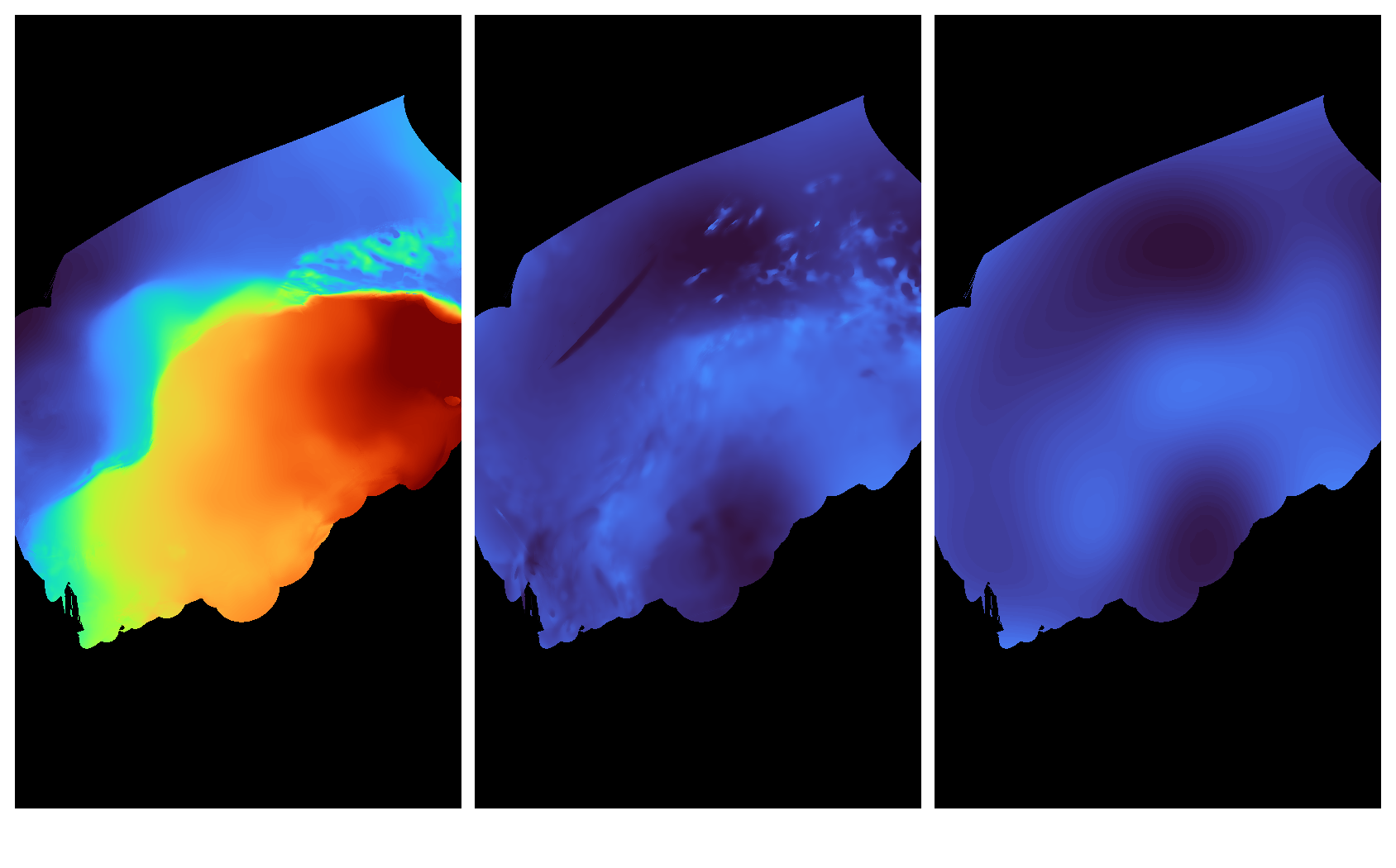}
    \caption{\textbf{Depth-supervised postoperative alignment of the case postop 1.}
    From left to right: initial, optimized, and target postoperative-MRI depths.}
    \label{fig:postop-depth}
\end{figure}

\section{Postoperative Surface Alignment}
\label{sec:postop}

\subsection{Problem setup and optimization strategy}
We use postoperative alignment as a representative scenario to test the geometric and rendering flexibilty of the proposed H-2DGS. In the postoperative scenario, the microscope reconstruction contains the post-resection photorealistic cortex, and the postoperative MRI supply information after the surgery. These two data modalities typically disagree geometrically, due to the resection cavity that significantly changes the surface morphology. We use the depth map as the proxy of the surface geometry to supervise the deformation. The inputs are an initial height-field model $\mathcal G_0$ reconstructed from microscope video and a depth dataset
\begin{equation}
    \mathcal D_{\mathrm{depth}}=\{(D_v,\Pi_v)\}_{v=1}^{V}
    \label{eq:postop-data}
\end{equation}
rendered from the postoperative MRI-derived mesh.
For the same cameras, the current height-field model produces differentiable depths $\hat D_v$.
The desired output is a deformed version of $\mathcal G_0$ whose surface follows the MRI depth while retaining the initial cortical texture.
This gives a direct image-space comparison without establishing sparse cross-modal point correspondences.

Naive depth fitting, however, admits undesirable solutions: neighboring patches can move independently and tear, opacity can change which surface is observed, and texture can smear while geometry moves.
Our optimization therefore combines three mechanisms with distinct roles.
Depth loss supplies the target shape, the transition objective couples adjacent charts, and tracked self-supervision transports the original appearance.
Alternating geometry and appearance windows keep the latter two goals from competing through the same updates.

\subsection{Depth-supervised surface deformation}

Centers, rotations, tangent scales, appearance coefficients, opacity, and height coefficients are trainable, allowing both global displacement and local reshaping.
Basis centers and bandwidths remain fixed so that every primitive retains a stable local coordinate system during deformation.
We reuse the explicit ray--surface intersection in \cref{eq:intersection} to obtain camera-linear depth.
For this pass, opacity is overridden to $0.999$; otherwise, the model could reduce the loss by making a mismatched foreground patch transparent rather than moving its geometry.
With rendered and target validity masks $M_v$ and $M_v^\star$,
\begin{equation}
    \mathcal L_{\mathrm{depth}}=
    \frac{\sum_{\mathbf p}M_vM_v^\star
    |\hat D_v(\mathbf p)-D_v(\mathbf p)|}
    {\sum_{\mathbf p}M_vM_v^\star+\epsilon}.
    \label{eq:depth-loss}
\end{equation}
Centers use an exponentially decayed learning rate from
$1.6{\times}10^{-3}$ to $1.6{\times}10^{-5}$; tangent scale uses $10^{-3}$, rotation and height coefficients use $10^{-4}$, and opacity uses $10^{-8}$.
This hierarchy makes center and scale the dominant deformation variables, uses rotation and height for smaller local corrections, and effectively stabilizes visibility.
The transition objective in \cref{eq:transition} remains active to prevent independently moving charts from tearing the surface.

\subsection{Tracked appearance preservation}

The MRI mesh constrains shape but contains no information about the color that should remain attached to each cortical location.
We therefore use the initial reconstruction as its own appearance reference.
Before deformation, valid pixels from reference renderings are back-projected and bound to local coordinates on candidate primitives.
Each tracked sample stores its initial rendered color:
\begin{equation}
    \mathcal Q=\{(i_m,u_m,v_m,h_m,\mathbf c_m^0)\}.
    \label{eq:tracked-set}
\end{equation}
where $i_m$ indexes the associated height-field primitive, $(u_m,v_m,h_m)$ are its local surface coordinates, and $\mathbf c_m^0\in\mathbb R^3$ is its initial rendered RGB color. 

As the primitives deform, samples move with their charts and are reprojected. We then encourage the reprojected color to be consistent with the stored initial color using a self-preservation term:
\begin{equation}
    \mathcal L_{\mathrm{track}}=
    \mathbb E_{m\in\mathcal Q}
    \left[
    \left\|
    \mathbf C_{\mathrm{cur}}(\pi_v(\mathbf x_m(\mathcal G)))
    -\mathbf c_m^0
    \right\|_1
    \right].
    \label{eq:track}
\end{equation}
Each sample is bound to eight candidate primitives, and targets are aggregated in image space. By enforcing the pixel-wise color consistency before and after deformation, the overall texture pattern of the surface is preserved. 

\subsection{Alternating geometry and appearance updates}

Applying depth and appearance updates simultaneously can yield an unstable compromise: the former must change geometry, while the latter resists changes to the rendered image.
We instead separate their dominant effects.
Optimization alternates a geometry-active window
\begin{equation}
    \mathcal L_{\mathrm{depth\text{-}win}}=
    \mathcal L_{\mathrm{depth}}+
    \lambda_{\mathrm{trans}}\mathcal L_{\mathrm{trans}}+
    \lambda_{\mathrm{track}}\mathcal L_{\mathrm{track}}
    \label{eq:depth-window}
\end{equation}
with an appearance window that freezes geometry and minimizes
$\mathcal L_{\mathrm{track}}$ using multi-view augmentation. We set $\lambda_{\mathrm{trans}}$ to 6.0 and $\lambda_{\mathrm{track}}$ to 10.0.
The first window moves and reshapes the surface under MRI supervision while maintaining local continuity and a weak appearance anchor.
The second cannot undo the acquired geometry; it only restores the texture fields around their deformed locations.
Together, the two windows separate MRI-driven shape changes from restoration of the original photorealistic appearance.

\section{Experiments}
\label{sec:experiments}

\subsection{Setup and metrics}

We evaluate five MRI cases using real neurosurgical videos: three preoperative cases and two postoperative cases. For each case, we select the continuous segment of the surgical video with the most comprehensive viewpoint coverage and uniformly sample frames from the original segment, resulting in a continuous sequence of 60--80 images for training.
The data were shared under the approved data-sharing agreement between the participating institutions.
All patients provided written informed consent for the use of their data for research purposes.
All experiments run on an NVIDIA RTX~3050 GPU with 4\,GB memory, 32\,GB system memory, and an Intel i7-12700H CPU.
\method uses 2,400 primitives per case, with 50 color and 20 opacity bases per primitive.
We compare against 2DGS~\cite{2DGS}, 3D-HGS~\cite{3d-hgs}, and QGS~\cite{QGS}, trained with their released default settings.

PSNR, SSIM, and LPIPS evaluate appearance.
PSNR and SSIM use the intersection of valid rendered pixels across methods.
Model size and per-frame time measure efficiency.
For geometry, multi-view depth consistency (MVDC) back-projects valid depth from one evaluation view, reprojects it into a neighboring view, and reports the median absolute depth discrepancy normalized by scene extent.

\begin{table*}[t]
\centering
\caption{Quantitative comparison on five surgical cases.
MVDC is normalized by scene extent.
Lower is better for LPIPS, MVDC, size, and time.
Missing LPIPS values were not reported by the corresponding baseline runs.}
\label{tab:quantitative}
\setlength{\tabcolsep}{3.3pt}
\renewcommand{\arraystretch}{0.64}
\resizebox{0.75\textwidth}{!}{
\begin{tabular}{clrrrrrr}
\toprule
Case & Method & PSNR$\uparrow$ & SSIM$\uparrow$ & LPIPS$\downarrow$ &
MVDC$\downarrow$ & Size (MB)$\downarrow$ & Time (ms)$\downarrow$\\
\midrule
Preop 1 & \method & 23.84 & 0.776 & 0.218 & \(\boldsymbol{4\mathrm{e}{-5}}\) & \textbf{7.97} & 39.13\\
  & 2DGS    & 28.54 & 0.901 & 0.109 & \(7\mathrm{e}{-5}\) & 40.87 & \textbf{25.42}\\
  & 3D-HGS  & 28.00 & 0.887 & --    & \(1.18\) & 32.24 & 152.81\\
  & QGS     & \textbf{31.46} & \textbf{0.952} & -- & \(9\mathrm{e}{-5}\) & 86.04 & 120.68\\
\midrule
Preop 2 & \method & 23.29 & 0.810 & 0.212 & \(\boldsymbol{2\mathrm{e}{-5}}\) & \textbf{7.97} & 29.68\\
  & 2DGS    & 27.35 & 0.906 & 0.109 & \(6\mathrm{e}{-5}\) & 45.72 & 29.69\\
  & 3D-HGS  & 27.46 & 0.900 & --    & \(1.57\) & 36.39 & 178.32\\
  & QGS     & \textbf{30.64} & \textbf{0.956} & -- & \(1\mathrm{e}{-4}\) & 107.03 & 164.09\\
\midrule
Preop 3 & \method & 25.20 & 0.839 & 0.161 & \(\boldsymbol{1\mathrm{e}{-5}}\) & \textbf{7.97} & 31.98\\
  & 2DGS    & 33.33 & 0.956 & 0.048 & \(3\mathrm{e}{-5}\) & 51.69 & 28.21\\
  & 3D-HGS  & 31.69 & 0.941 & --    & \(1.41\) & 36.88 & \textbf{6.16}\\
  & QGS     & \textbf{40.07} & \textbf{0.986} & -- & \(5\mathrm{e}{-5}\) & 93.64 & 152.35\\
\midrule
Postop 1 & \method & 21.54 & 0.745 & 0.210 & \(\boldsymbol{4\mathrm{e}{-5}}\) & \textbf{7.97} & 40.13\\
  & 2DGS    & 29.23 & 0.915 & 0.075 & \(1\mathrm{e}{-4}\) & 70.14 & 33.38\\
  & 3D-HGS  & 28.26 & 0.895 & --    & \(1.0\) & 51.95 & \textbf{8.94}\\
  & QGS     & \textbf{32.03} & \textbf{0.964} & -- & \(1.4\mathrm{e}{-4}\) & 142.01 & 188.04\\
\midrule
Postop 2 & \method & 23.53 & 0.797 & 0.180 & \(\boldsymbol{3.2\mathrm{e}{-5}}\) & \textbf{7.97} & 42.47\\
  & 2DGS    & 30.56 & 0.933 & \textbf{0.073} & \(9.58\mathrm{e}{-5}\) & 46.26 & 24.78\\
  & 3D-HGS  & 29.51 & 0.921 & -- & \(1.42\) & 40.78 & \textbf{7.52}\\
  & QGS     & \textbf{33.92} & \textbf{0.968} & -- & \(1.18\mathrm{e}{-4}\) & 78.93 & 135.80\\
\bottomrule
\end{tabular}}
\end{table*}

\begin{figure}[!t]
    \centering
    \includegraphics[width=1.0\linewidth]{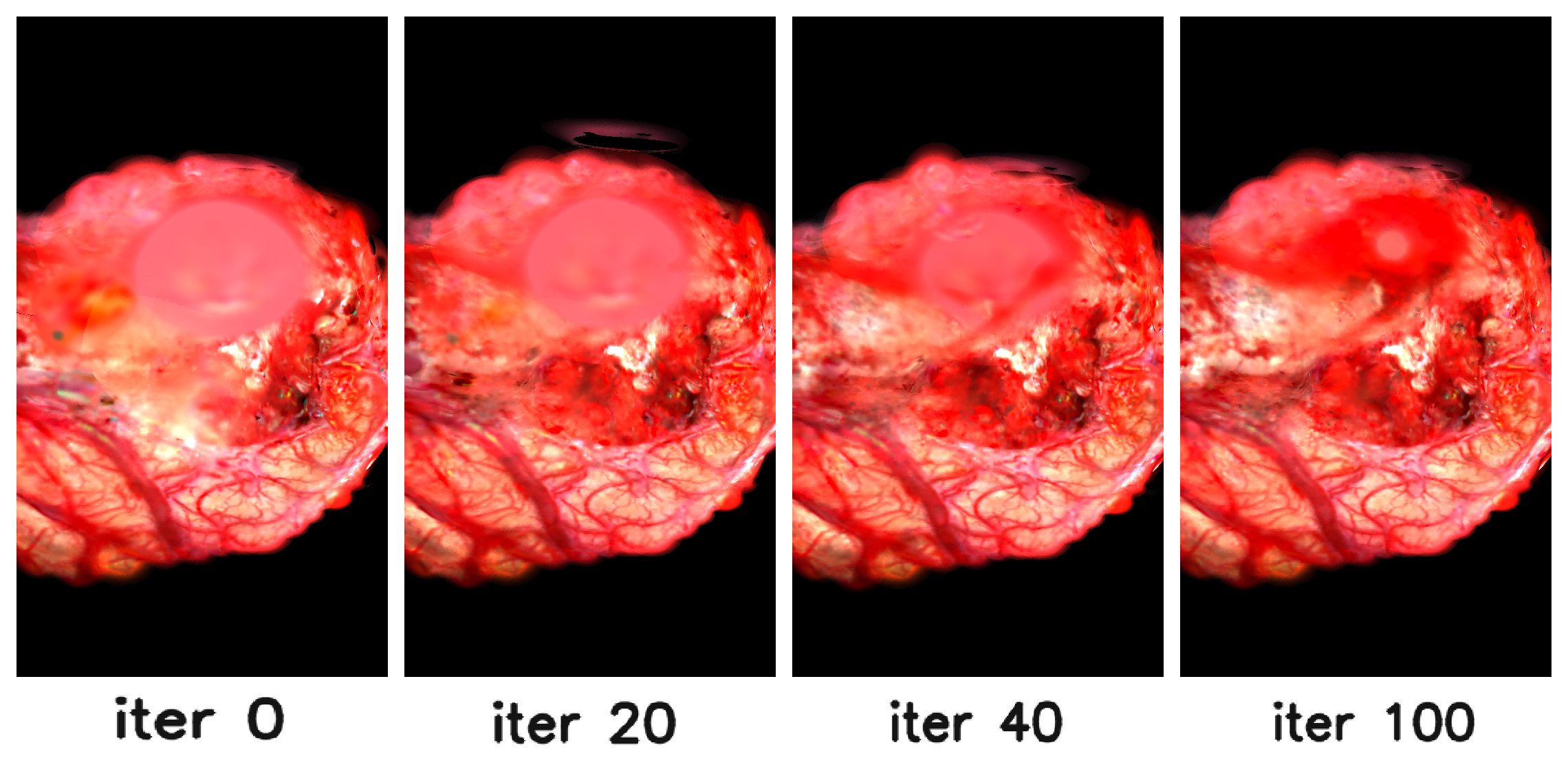}\\[1pt]
    \includegraphics[width=1.0\linewidth]{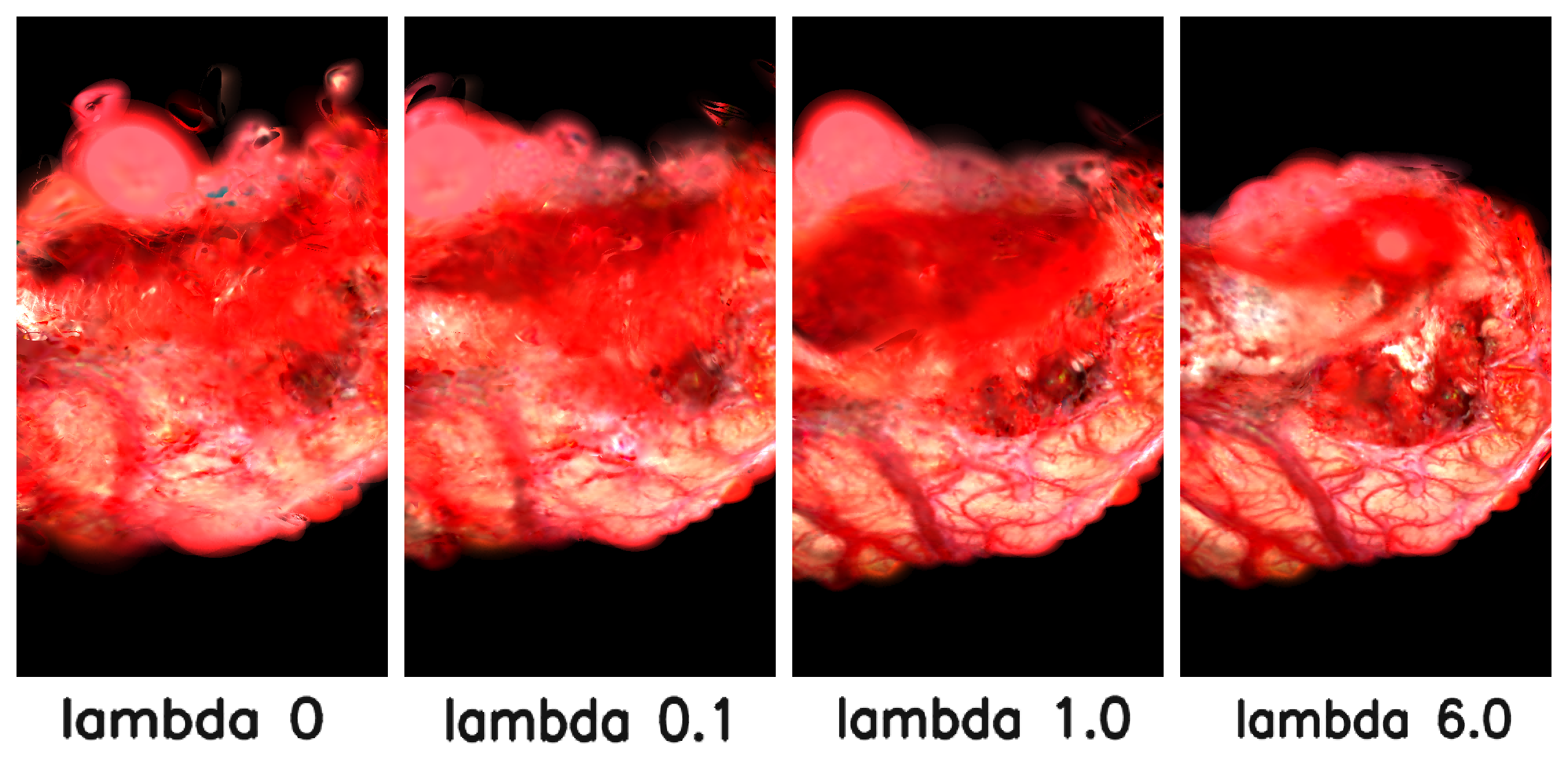}
    \caption{Components of postoperative deformation, case postop 1.
    Top: appearance-window iterations recover texture after depth fitting.
    Bottom: stronger geometry-continuity weighting coordinates neighboring patches.}
    \label{fig:postop-ablation}
\end{figure}

\subsection{Sharpness, consistency, and fusion}

\Cref{tab:quantitative} exposes the intended trade-off.
\method has lower training-view PSNR and SSIM than the baselines, mainly losing high-frequency highlights.
Its geometric behavior is substantially more stable: MVDC lies between
$1{\times}10^{-5}$ and $4{\times}10^{-5}$.
The representation is consistently 8\,MB and renders in 30--42\,ms.
QGS attains the strongest photometric scores but uses 79--142\,MB and 120--188\,ms.

Differences that are muted in covered views become pronounced away from the training trajectory.
\Cref{fig:uncovered} shows that volumetric or weakly constrained representations can extend into streaks and fragments.
\method remains a recognizable cortical layer.
This matters for anatomical fusion: in \cref{fig:fusion-comparison}, fibers appear attached to a defined surface instead of intersecting a diffuse cloud.
The tractography therefore serves both as a target application and as a visual stress test of geometric localization~\cite{chamberland2014realtime}.

\subsection{Postoperative deformation}

On the resection case, depth fitting moves the post-resection reconstruction toward postoperative MRI-rendered depth.
The optimized map follows the target, while the surface stretches laterally and the cavity becomes shallower without losing major tissue or vessel patterns (\cref{fig:postop-before-after,fig:postop-depth}).

Depth supervision alone destroys high-frequency appearance.
Appearance-window updates restore cortical texture with geometry frozen, while stronger transition regularization improves local geometric and texture coherence (\cref{fig:postop-ablation}).
The final overlay places the photorealistic cavity at the corresponding postoperative MRI region across viewpoints (\cref{fig:teaser-postop}).

\section{Discussion and Conclusion}
\label{sec:conclusion}

Together with the proposed three-stage pipeline, \method produces a photorealistic reconstruction of the cortical surface with a sharp geometry and well-defined boundary. This explicit surface enables direct integration with subcortical tractography and whole-brain MRI context without the internal anatomy being obscured by diffuse volumetric opacity. Compared with the baseline approaches, \method achieves substantially better geometric generalization across viewpoints: it consistently maintains a localized cortical surface in uncovered views and obtains lower multi-view depth inconsistency across the five cases (\cref{tab:quantitative,fig:uncovered}).

The clear geometry of \method also provides a suitable foundation for deformation. Its surface primitives retain sufficient geometric flexibility to represent substantial morphological changes, allowing deformation objectives to be defined directly on the primitive parameters and fitted to a target geometry. The geometric and rendering flexibility of H-2DGS is verified in the postoperative case, where the fitting process is separated into depth-fitting and appearance-preservation stages. The postoperative experiment demonstrates that this strategy can accommodate a pronounced geometric discrepancy while preserving the photorealistic appearance of the original reconstruction.

The compact size of \method follows from its fixed 2,400-primitive budget with per-primitive texture on an explicit surface. We do not match primitive counts to the baselines, since volumetric primitives lacking local texture generally need more of them to fit the same images. Overall, the smaller footprint reflects an advantage of the height-field primitive, especially given its stronger geometric localization.

A remaining limitation is rendering quality. As indicated by PSNR and SSIM, \method does not reproduce the observed images as accurately as the baseline methods. Replacing volumetric support with a thin explicit surface removes geometric ambiguity, but also reduces the rendering expressiveness available for reproducing high-frequency appearance variations. We introduce spatially varying texture through the color bases to compensate for this loss, but a clear photometric gap remains. Future work should therefore investigate richer geometry-preserving appearance models that improve rendering fidelity without reintroducing diffuse volumetric opacity.

The current pipeline also depends on camera poses recovered by structure from motion (SfM), whose static-scene assumption can be violated in surgical videos. One additional sequence failed because physiological cortical pulsation introduced non-rigid inter-frame motion, causing SfM to confound tissue motion with camera motion and produce inaccurate poses. This error subsequently degraded all reconstruction stages. Potential solutions include obtaining calibrated or externally tracked camera poses from specialized surgical equipment, representing each primitive's geometric attributes as continuous functions of time, and estimating the dominant pulsation frequency for phase-aware frame selection or motion correction. Addressing both appearance expressiveness and dynamic tissue motion will be important for extending \method to more diverse surgical videos. In the future, combining this approach with cinematic rendering~\cite{comaniciu2016shaping,kraaijeveld2026real} should offer a unique opportunity to improve surgical navigation.

{\small
\bibliographystyle{ieeenat_fullname}
\bibliography{main}
}

\end{document}